\documentstyle[12pt]{article}

\input{epsf.tex}

\title{Testing the Cactus code on exact solutions of the
Einstein field equations}

\author{Dumitru N. Vulcanov\thanks{
Permanent address : The West University of Timi\c soara,
Theoretical and Computational Physics Department,
B-dul V. P\^ arvan no. 4, 1900 Timi\c soara,  Rom\^ ania, e-mail : 
{\tt vulcan@physics.uvt.ro}}
\/ and Miguel Alcubierre\thanks{e-mail : {\tt miguel@aei.mpg.de}}\\
Max-Planck-Institut f\" ur Gravitationsphysik\\
Albert-Einstein-Institut\\
Numerical Relativity Group\\
Golm, Am M\" uhlenberg 1, D-14476, Germany}

\begin{document}

\date{}
\maketitle

\begin{abstract}
  The article presents  a series of numerical simulations of exact solutions of
  the Einstein equations performed using the Cactus code, a complete
  3-dimensional machinery for numerical relativity.  We describe an
  application (``thorn'') for the Cactus code that can be used for
  evolving a variety of exact solutions, with and without matter,
  including solutions used in modern cosmology for modeling the early
  stages of the universe. Our main purpose has been to test the Cactus
  code on these well-known examples, focusing mainly on the stability
  and convergence of the code.
\end{abstract}

\section{Introduction}

Numerical Relativity is concerned with the study of numerical
solutions of the Einstein's equations for the gravitational field,
which are at the core of the theory of General Relativity. General
Relativity (\cite{1}-\cite{2}) is a 4-dimensional theory involving one
dimension of time and three of space. The field equations, called
Einstein equations, are :
\begin{equation}\label{EE}
R_{ij} - \frac{1}{2}g_{ij} R + \lambda g_{ij} = \frac{8 \pi G}
{c^4} T_{ij} \, ,
\end{equation}
where $\lambda$ is the cosmological constant, $R_{ij}$ the Ricci tensor, 
$R$ the Ricci scalar, $g_{ij}$ the spacetime metric, $T_{ij}$ the 
stress-energy tensor, G the gravitational constant, $c$ the speed of light and
 $i,j=0,1,2,3$.

These equations are an
extremely complicated system of coupled, non-linear, partial
differential equations and solving them numerically makes enormous demands on
the processing power and memory of a computer. Because of this,
Numerical Relativity (\cite{4}, \cite{5}) has proceeded in several
stages, first solving 1-dimensional problems (that is 1 spatial
dimension, e.g. spherical symmetry), and then moving to 2-dimensional
problems (i.e.  axial symmetry).  Only in the last few years have
computers developed sufficiently to consider tackling fully
3-dimensional problems. At the Albert Einstein Institute, 
an international team
led by  Edward Seidel has developed a complete 3D code for
numerical relativity, which has been named the ``Cactus code'' (see
\cite{6}).  The Cactus code has been mainly
designed as a computational toolkit (freely available for the
scientific community) for simulating different systems of partial
differential equations.  In the particular of relativity, the code can
be used to simulate systems with strong gravitational fields:
collapsing gravitational waves, colliding black-holes, neutron stars,
and other violent astrophysical processes generating gravitational
waves \cite{7}-\cite{13}.

Among the main problems related to the development of the Cactus code
has been that of testing the code for stability and convergence during
the simulations. Several applications of the code (named generically
``thorns'') were designed for this purpose, one of them being the
so-called thorn ``Exact'', were some known exact solutions of the
vacuum Einstein equations  were implemented. The original version
of this thorn was written and has
been furthered developed by many people.  Thorn Exact was designed for
a comparative study of the numerical evolution of exact solutions of
the vacuum Einstein equations~\cite{7}.  The thorn requires the full
4-dimensional metric of a given exact solution in the whole spacetime.
Its routines then generate 3+1 data (i.e. lapse function, shift
vector, spatial metric, and extrinsic curvature) from the given exact
solution both as initial data for the numerical evolution, and also at
every iteration step of the evolution so that it can be directly
compared with the numerically evolved data.  A limited number of exact
solutions were included (among them Minkowski, Novikov and several
Black Hole spacetimes), but since the thorn was constructed in very
modular form it has been easy to add new exact solutions by just
writing a special subroutine for each new case (with the corresponding
parameters). Thus we have added a series of space-times which are of
cosmological interest, such as De Sitter, Friedmann, Kasner,
DeMilne and G\" odel models. The main problem was that some of these
models are exact solutions of the Einstein equations with matter, 
so the original thorn has
to be generalized to allow one to introduce the components of the
corresponding stress-energy tensor.

The current version of this thorn Exact now contains also several 
space-time models used in cosmology for treating the early stages 
of the universe. Thus, through this thorn, the Cactus code
can be used for numerical cosmology studies, an example being the
inflationary cosmology where inflation is controlled by a scalar field
(this is the main purpose of our future work).

This article is a report on the current status of thorn Exact. We
shall present some of the numerical results obtained on running the
Cactus code for several cosmological models introduced in this thorn.
We point out here also how the code produces convergent numerical
results and how stable the evolution can be for different cases. At
the moment the Cactus code can use two different formulations for
numerical integrating the Einstein equations , namely the standard
Arnowitt-Deser-Misner formulation (\cite{1},\cite{3}) (implemented as
the ADM thorn) and a recent reformulation introduced by Baumgarte and
Shapiro~(\cite{11}, \cite{14}) based on previous work of Shibata and
Nakamura~(\cite{15}) (implemented in the BSSN
thorn). The comparative study of these two formulations shows the
better stability properties of the BSSN formulation (as already
pointed out in \cite{11}).  A special point to
mention here is the problem of the boundary behavior of the numerical
simulations. The boundary conditions currently available in the Cactus
code are specifically  adapted for the simulations of either
asymptotically flat spacetimes, or periodic spacetimes.  Because of
this we have been forced to use special boundary conditions in thorn
Exact which force the code to take the boundary values from the exact
solution itself, except in those cases were we can use the ``flat''
boundary condition already implemented in the code (the flat boundary
condition means that the value of the given field or fields at the
boundary is simply copied from the value one grid point in along the
direction normal to the boundary).

In the next sections we shall present several results obtained after
running Cactus for different metrics given by thorn Exact, on
different computer architectures. The simulations were performed
mainly on single processor machines, using both the AEI computer
network (SGI or Dec machines with UNIX operating system) and a Pentium
III machine with a UNIX FreeBSD operating system at the West University of 
Timi\c soara.  Some of the simulations, with similar results, were also done
on the Origin 2000 supercomputer at the AEI. As a conclusion of our
work we have extended the Exact thorn to deal with exact solutions of
the Einstein equations  with matter, in particular several cosmological solutions.  The thorn is available, via internet downloading, for anyone interested 
in using  it.

Through this article and in the  Cactus code we shall use geometrical units 
with $G=c=1$.


\section{De Sitter cosmologies}

Modern cosmology is based on the Robertson-Walker metric (\cite{16},\cite{17}) 
having the line element as :
\begin{equation}\label{RW}
ds^2 = -dt^2 + R(t)^2 \left [ \frac{dr^2}{1-kr^2} + r^2 (d\theta^2 +
\sin^2 \theta~d\phi^2)\right ] \nonumber
\end{equation}
in spherical coordinates $(t,r,\theta,\phi)$. Here $R(t)$, the scale
factor, is depending only on the time $t$. The parameter $k$ has three values,
namely $k = 0, 1, -1$ for flat, closed or respectively open spatial three
geometry. The Einstein equations (\ref{EE})
for $\lambda=0$ and a stress-energy tensor as for a perfect fluid 
\begin{eqnarray}
T_{ij} = (p + \rho)u_i u_j + p g_{ij} \nonumber
\end{eqnarray}
(where $p$ is the pressure, $\rho$ the density and $u_i$ the velocity in
comoving coordinates) have, as an exact solution, for a
matter dominated universe ($p=0$) and flat spatial geometry ($k=0$) 
the DeSitter metric (also called Einstein-DeSitter metric, 
see \cite{1}, \cite{2}, \cite{16} and
\cite{17}) where $R(t) \sim t^{2/3}$. Thus DeSitter metric has the line
element as :
\begin{equation}\label{Desitter_metric}
ds^2 = -dt^2 + at^{4/3}\left [ dr^2 + r^2\left (d\theta^2 + 
\sin(\theta)^2 d\phi^2\right )\right] \, .
\end{equation} 
where $a$ is a constant. Here the only non-vanishing component of the 
the stress-energy tensor is :
\begin{eqnarray}
T_{00} = \frac{1}{6\pi t^2} \, .
\end{eqnarray}

\begin{figure}
\epsfxsize=3.0in
\epsfysize=3.0in
\epsfbox{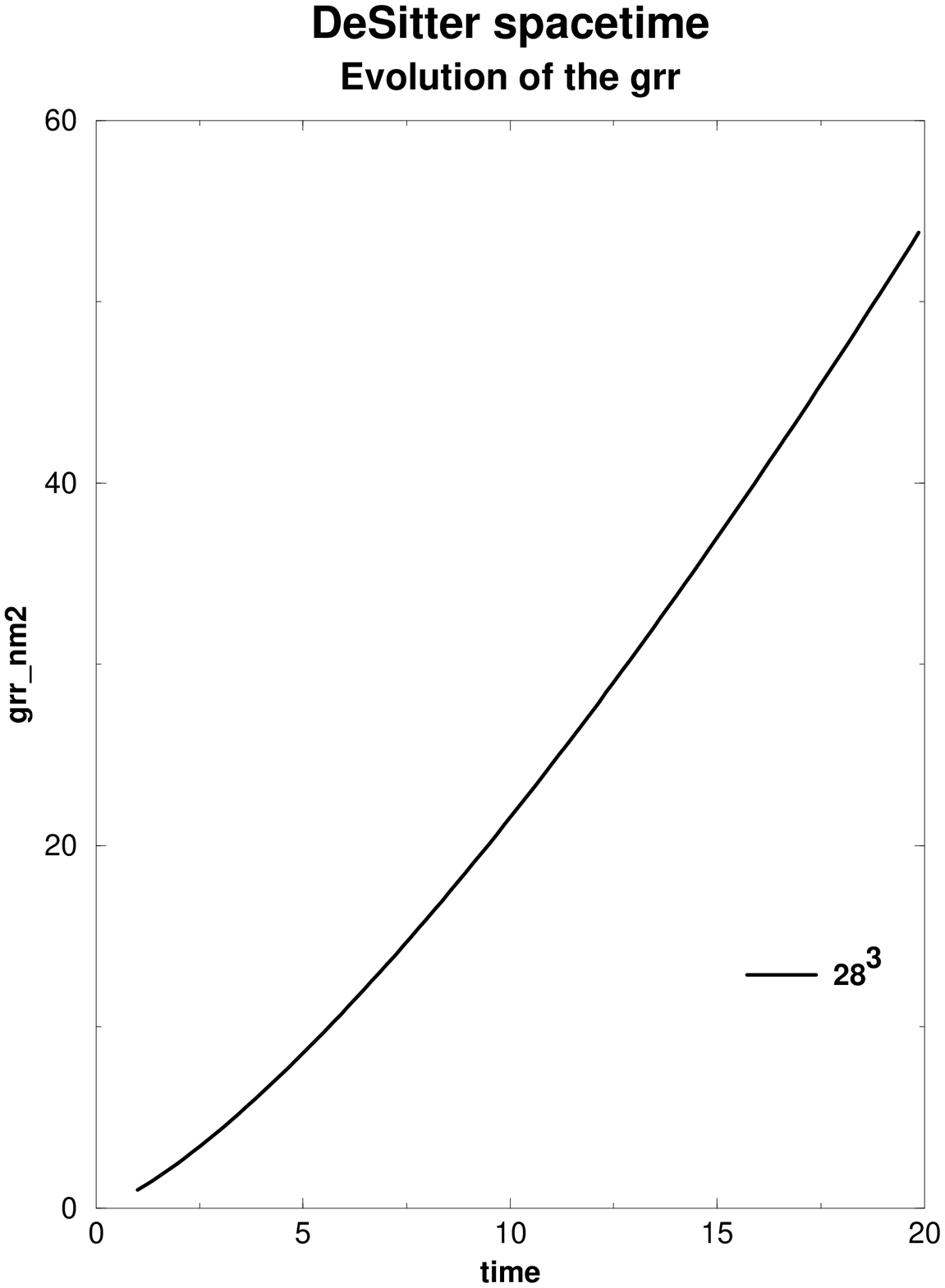}
\vspace{-3.0in}
\hspace{3.2in}
\epsfxsize=3.0in
\epsfysize=3.0in
\epsfbox{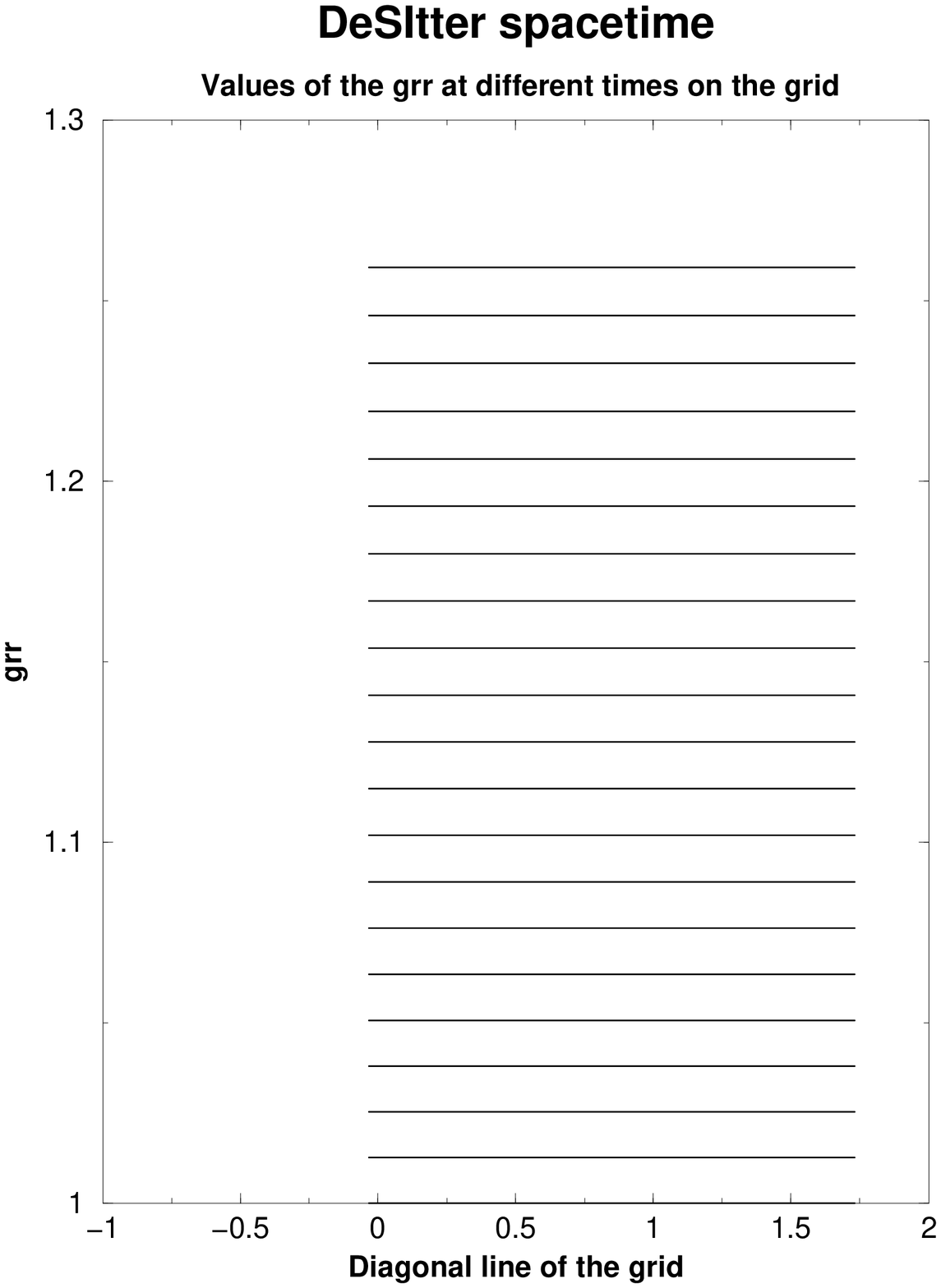}
\caption{Evolution of the L2 norm of the radial metric function $g_{rr}$
  versus time (left panel) and of the metric function itself
  through the grid at different times (right panel) for the De Sitter
  spacetime.  The numerical grid has $28^3$ points.}
\label{fig:desitt_grr}
\end{figure}

Next we shall present some of the results obtained after running
Cactus for the De Sitter metric. First we show the behavior of the
radial metric function ($g_{rr}=R(t)^1/(1-k r^2)$  in the above line element, 
being denoted with {\bf grr} at the Cactus output)
obtained after 20 iterations on a grid of
$28^3$ points, with $0 \leq x,y,z \leq 1$ and $\Delta t = 0.25 \Delta x$
(Fig.~\ref{fig:desitt_grr}). Here and in the next figures,
``normalized'' means the L2 norm of the respective function calculated
using all the values of the function on the computational grid at one
time and the respective output Cactus files are denoted with {\bf \_nm2}
extension.

Cactus code is evolving the Einstein equations using the 3+1 decomposition
of space-time (both in ADM or BSSN evolution methods). Thus the Einstein
equations can be split in two groups : the dynamic equations (for the
time derivatives of the three-dimensional metric and extrinsic curvature)
and the constraint equations (the Hamiltonian constraint and the
Momentum constraint) - see \cite{1} and \cite{3}.  The constraint equations
 are  satisfied during all time evolution
of the system. Thus one of the main tests on the convergence in the 
Cactus code is given by the time behavior of the Hamiltonian constraint
(an output of the Cactus code through a thorn called ADMConstraints, 
namely {\bf ham}). In the next Figure ~\ref{fig:desitt_ham}
we show the convergence of the L2 norm of the Hamiltonian
constraint for DeSitter metric,  for 20 and 2000 iterations, using
two different resolutions on the grid, one with $14^3$ and the second
with $28^3$ points (both grids cover the same region of spacetime, so
the grid with more points has a smaller value of the finite difference
interval $\Delta x$).  Notice that the Hamiltonian constraint for a
true solution of the Einstein equations  should be equal to zero.  
Finite differencing errors imply that the numerical solution will have 
a non-vanishing value of the Hamiltonian constraint.  For a consistent finite
difference approximation of the Einstein equations, 
we should expect the Hamiltonian
constraint to approach zero as the resolution is increased.  For a
second order approximation, the value of the Hamiltonian constraint
should go down by a factor of four when the resolution is doubled.
The figure shows that we have close to second order convergence.

This was the method to test the convergence of the code in all examples
presented in this article. We have obtained good second order convergence
in all examples as is shown in the next figures nr ~\ref{fig:cosdesiter_ham},
~\ref{fig:kasner066}, ~\ref{fig:kasner-1}, ~\ref{fig:kasner-axi} and
~\ref{fig:kasner-gen}.


\begin{figure}
\epsfxsize=3.0in
\epsfysize=3.0in
\epsfbox{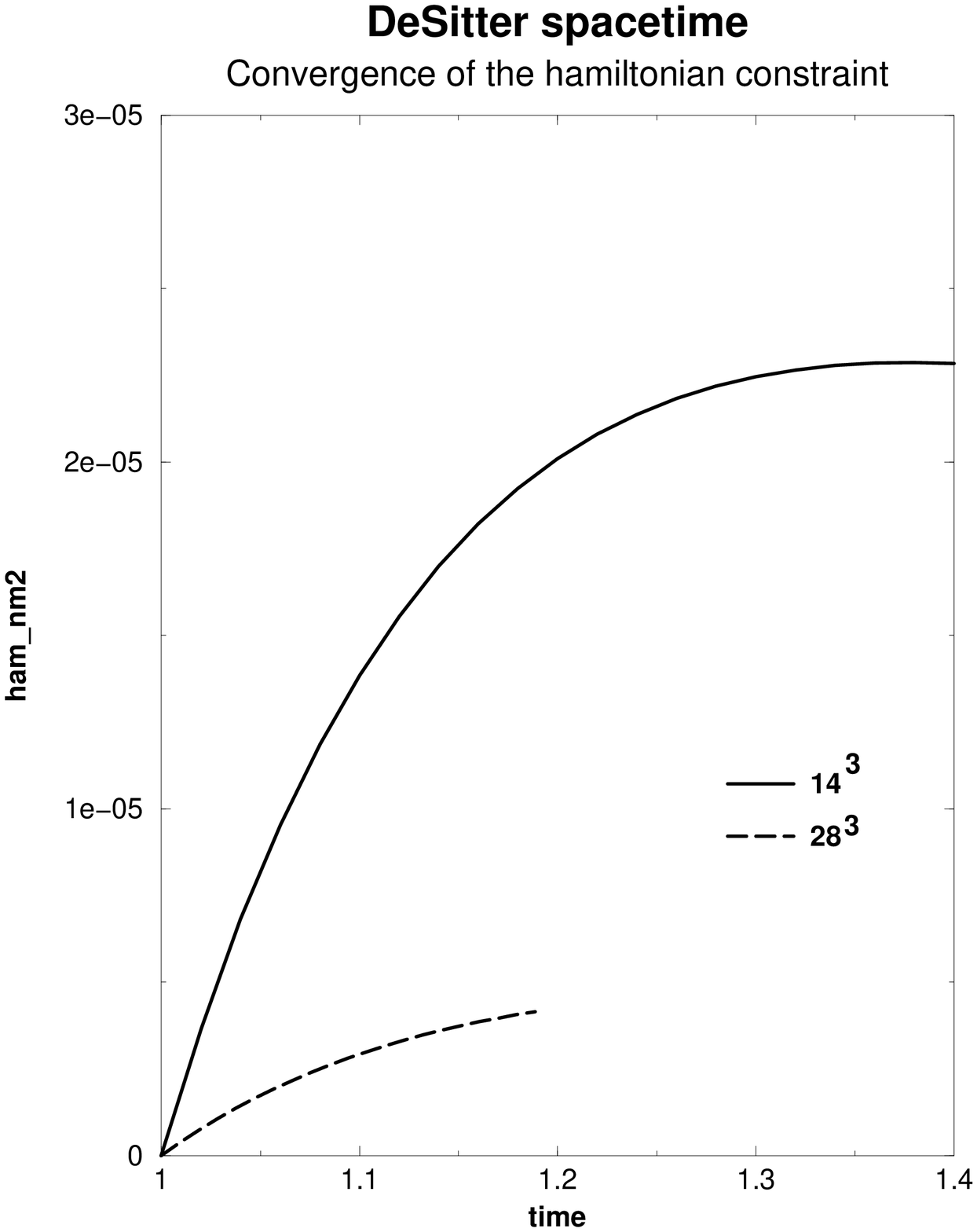}
\vspace{-3.0in}
\hspace{3.2in}
\epsfxsize=3.0in
\epsfysize=3.0in
\epsfbox{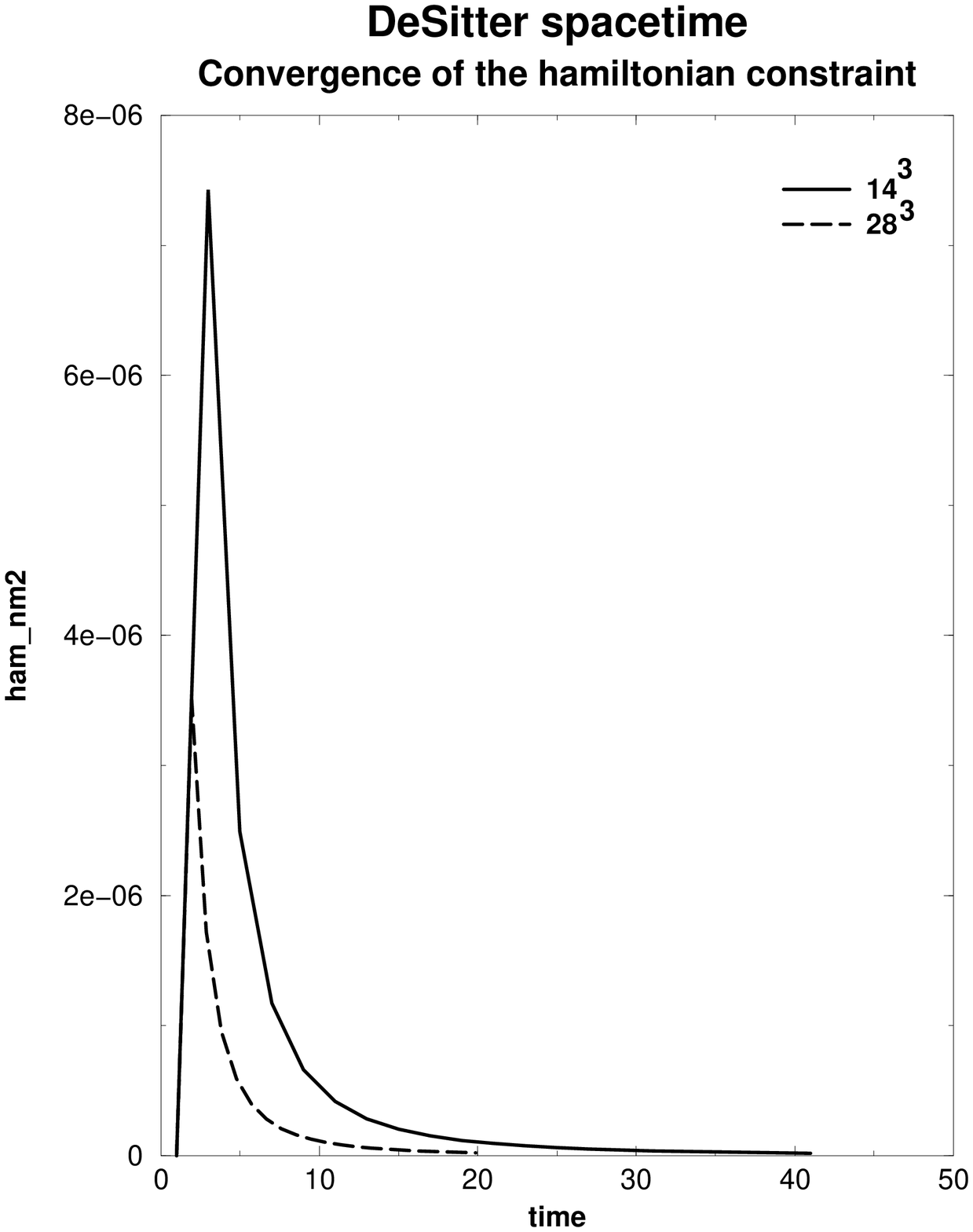}
\caption{Convergence of the L2 norm of the Hamiltonian constraint for
  the De~Sitter spacetime using two resolutions, one with $14^3$ and
  the second with $28^3$ points on the grid, for 20 iterations (left
  panel) and after 2000 iterations (right panel).}
\label{fig:desitt_ham}
\end{figure}

These results were obtained using the BSSN formulation for the
numerical evolution, but we have obtained similar results using the
ADM formulation. As a major conclusion we can point out that we 
found that the BSSN formulation results in stable evolutions. For long
time  evolutions (that is for more iterations than 20, for example 2000
iterations in our simulations), the Hamiltonian constraint grows from zero to a
constant value, as it is obvious by inspecting all our next graphs 
on the convergence of the Hamiltonian constraint (see below).
In our evolutions, one can also observe how the computational errors
propagate back into the computational grid after being reflected on
the boundary due to the use of inappropriate boundary conditions (here
we used the ``flat'' boundary conditions implemented in the Cactus
code). This will be an important problem to solve in the future.  The
De~Sitter parameter used in these simulations was $a=1.0$ and the
chosen slicing condition was an ``exact'' slicing (i.e. use the exact
lapse coming from Exact thorn).

\begin{figure}
\epsfxsize=3.0in
\epsfysize=3.0in
\epsfbox{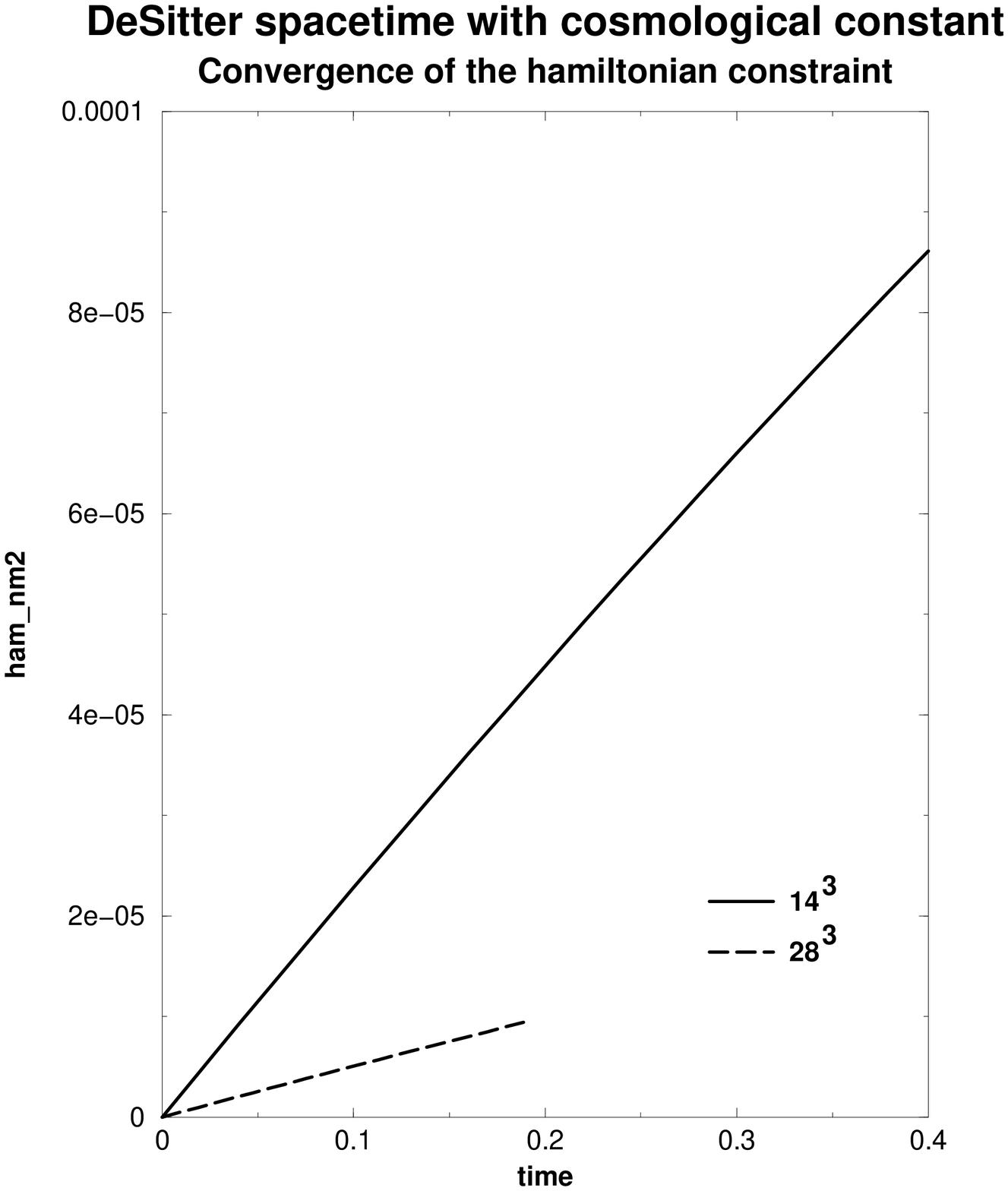}
\vspace{-3.0in}
\hspace{3.2in}
\epsfxsize=3.0in
\epsfysize=3.0in
\epsfbox{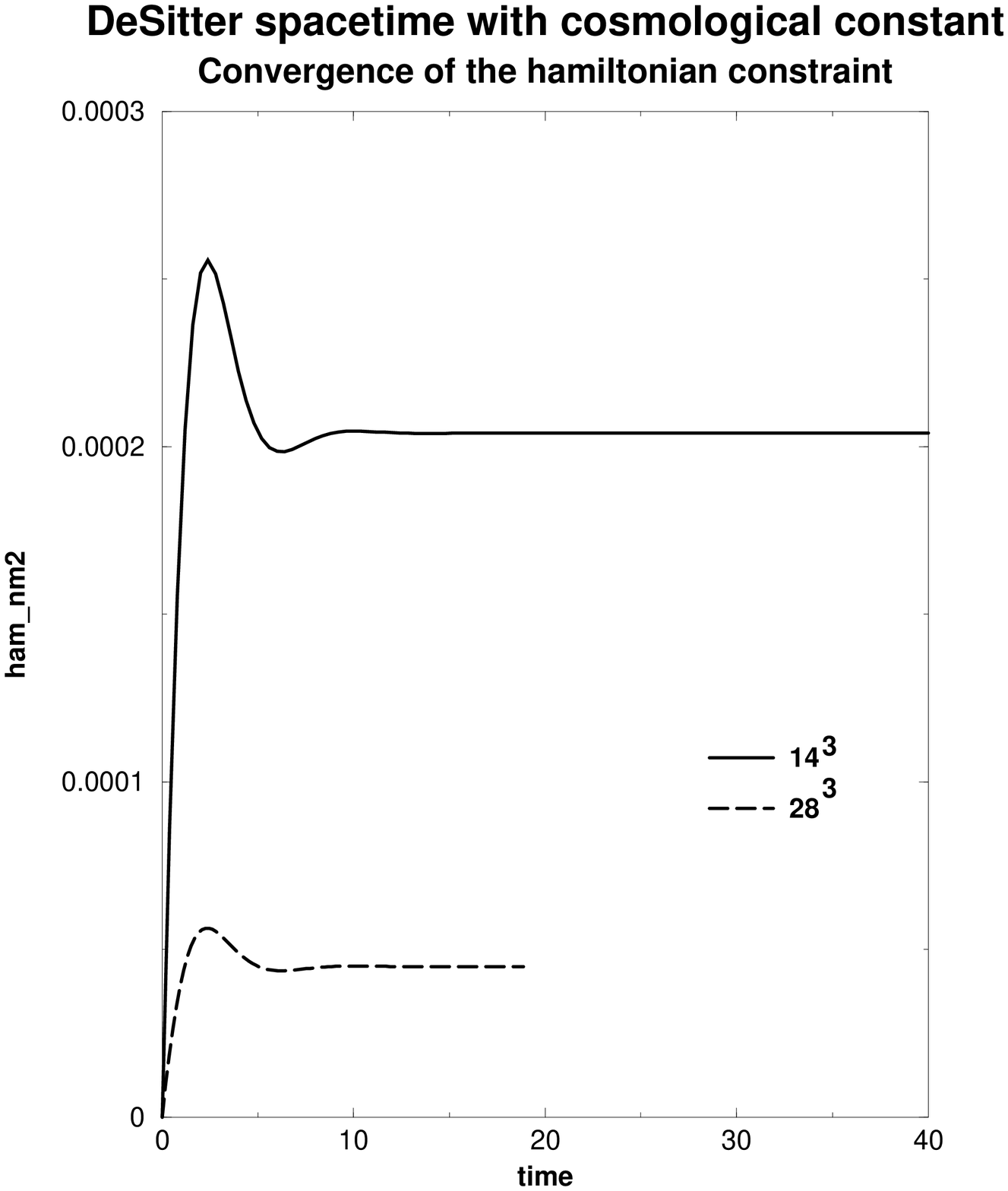}
\caption{Convergence of the Hamiltonian constraint for the De~Sitter
  spacetime with cosmological constant using two resolutions, one with
  $14^3$ and the second with $28^3$ points on the grid, for 20
  iterations (left panel) and after 2000 iterations (right panel).}
\label{fig:cosdesiter_ham}
\end{figure}

Another example of De Sitter cosmology is that of the De Sitter metric
with cosmological constant, having the form :
\begin{equation}\label{Cosdestter}
ds^2 = - dt^2 + e^{2/3\sqrt{3\lambda}t} \left ( dx^2 + dy^2 + dz^2
\right) \, .
\end{equation}
where $\lambda$ is the cosmological constant.
This is an example of an exact solution of the Einstein equations 
(\ref{EE}) with  cosmological constant and without matter. Here we can
still use the Cactus code, even though it has been written for the
case of no cosmological constant, by using a simple trick: we have
transferred the term with the cosmological constant to the right hand
side of the Einstein equations. Thus we shall have a non-vanishing
``matter'' term having as components
\begin{eqnarray}
T_{ij}= - \frac{\lambda}{8 \pi} g_{ij} = \left ( \begin{array}{cccc}
\frac{1}{8}\frac{\lambda}{\pi} & 0 & 0 & 0\\
0 & -\frac{1}{8}\frac{\lambda  e^{2/3 \sqrt{3\lambda}t}}{\pi }& 0 & 0\\
0 & 0 &-\frac{1}{8}\frac{\lambda e^{2/3 \sqrt{3\lambda}t}}{\pi }&  0\\
0 & 0 & 0 & -\frac{1}{8}\frac{\lambda  e^{2/3 
\sqrt{3\lambda}t}}{\pi}\end{array}\right ) \, .
\end{eqnarray}
(remember that we have $G=c=1$).
Introducing this in the Cactus code and taking $\lambda =1$,   we have
obtained the numerical results presented in the next figures 
~\ref{fig:cosdesiter_ham} and ~\ref{fig:cosdesitt_grr}. First,
we investigated the convergence of the code, using again the time
behavior of the Hamiltonian constraint.
Figure~\ref{fig:cosdesiter_ham} shows this for simulations with 20 and
2000 iterations, using again two resolutions, one with $14^3$ and the
second with $28^3$ points on the grid, with $0 \leq x,y,z \leq 6$ and
$\Delta t = 0.25$.  We can point out the same conclusions on the
convergence and the stability of the code as in the case of the
De~Sitter spacetime.

Figure~\ref{fig:cosdesitt_grr} presents the evolution in time of the
radial metric component $g_{rr}$ for 2000 iterations. It is easy to
see here the rapid increase of the radius of the universe (which is
related to $g_{rr}$), showing the inflationary nature of a universe
modeled by this metric.

\begin{figure}
\epsfxsize=3.0in
\epsfysize=3.0in
\epsfbox{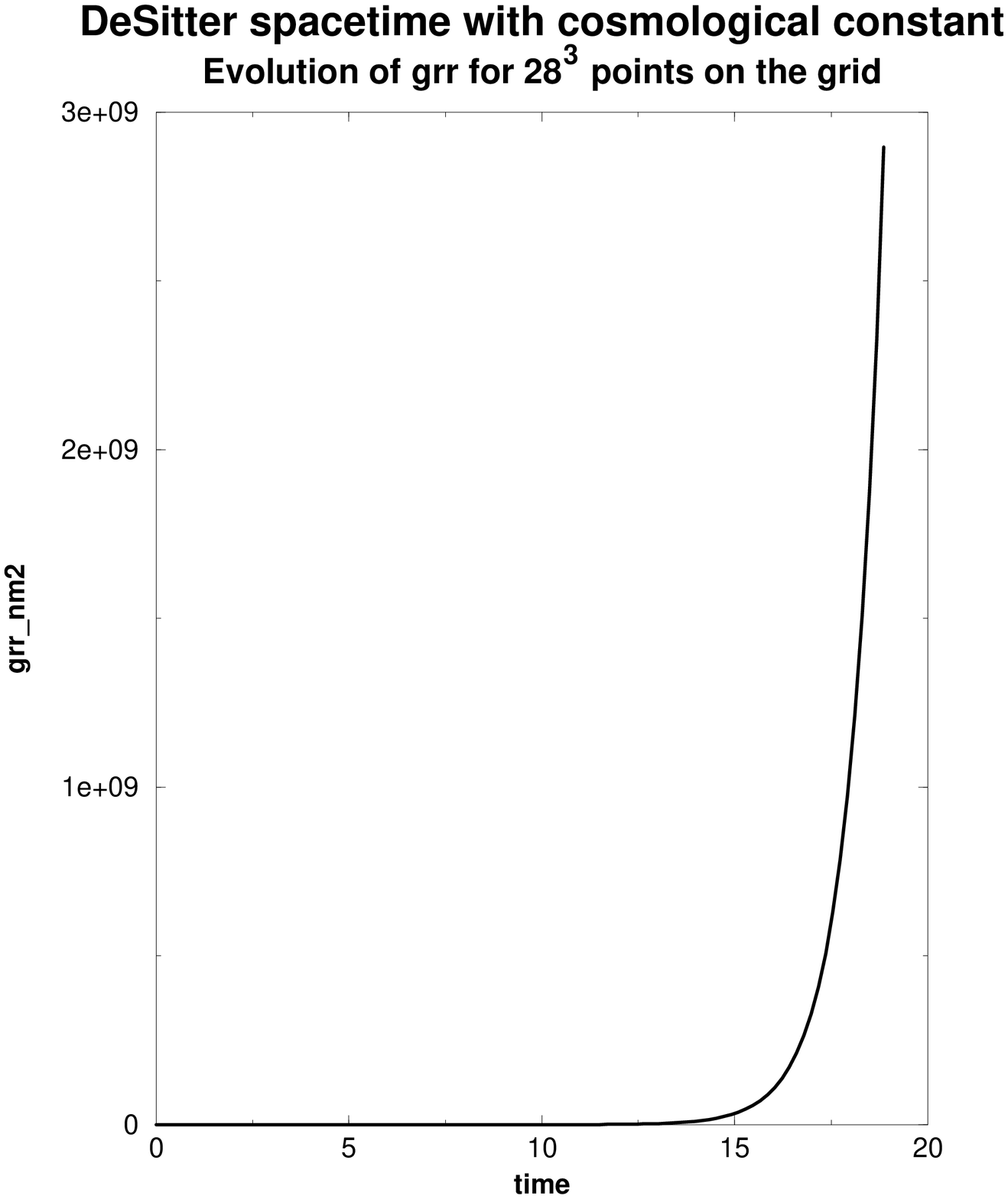}
\vspace{-3.0in}
\hspace{3.2in}
\epsfxsize=3.0in
\epsfysize=3.0in
\epsfbox{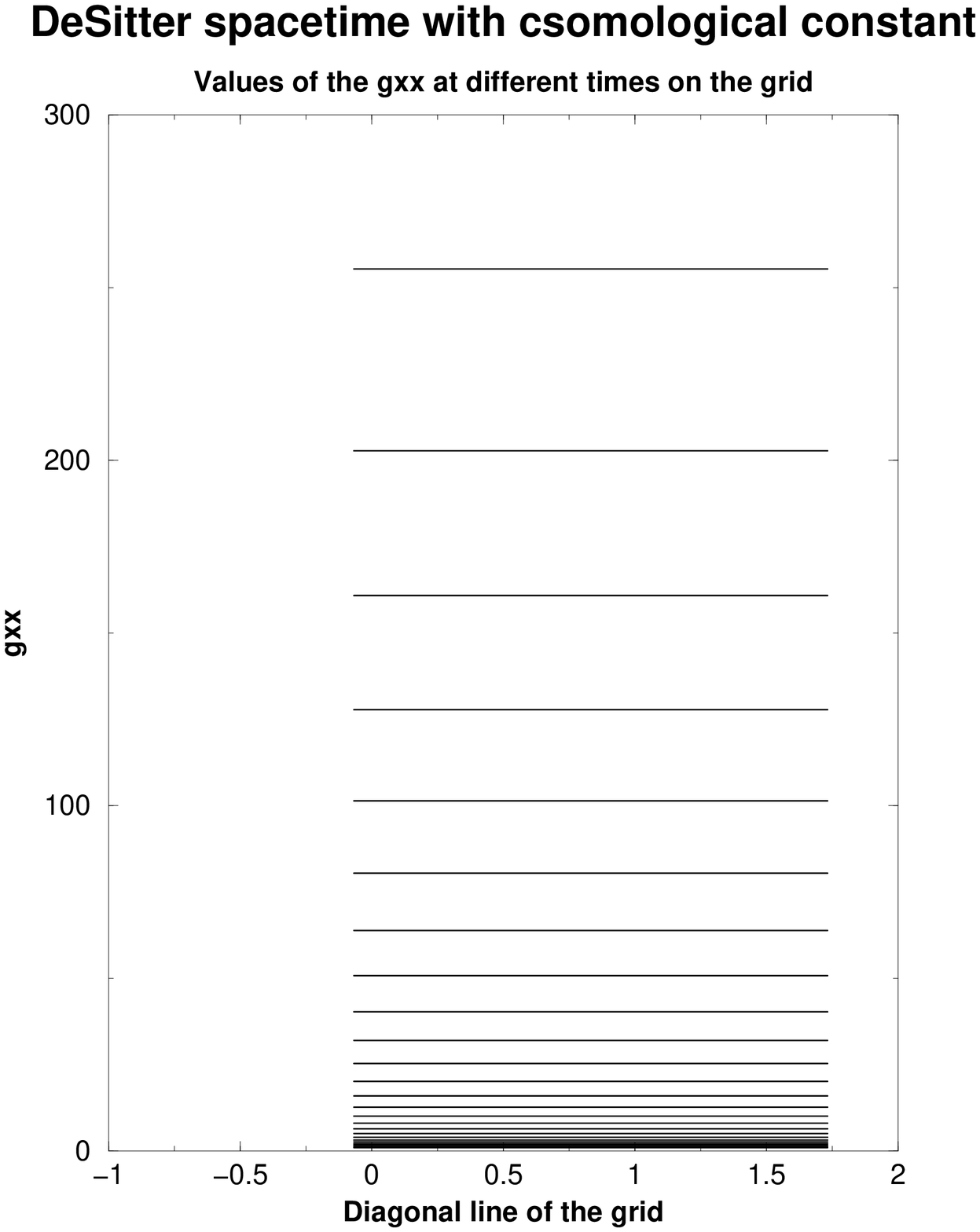}
\caption{Behavior of the L2 norm of the radial metric component $g_{rr}$
  (left panel) and its evolution in time (right panel) showing the
  inflationary nature of a universe modeled with the De~Sitter metric
  with cosmological constant.}
\label{fig:cosdesitt_grr}
\end{figure}

\section{Kasner solutions}

We have used the Cactus code for several cosmological solutions of the
Einstein equations, generically named ``Kasner'' metrics, included in
the class of homogeneous anisotropic models, which can emulate some
early stages of the Universe. One of these metrics is the so-called
``Kasner-like'' metric (See \cite{17} and \cite{19}), which in Cartesian
coordinates has the form:
\begin{eqnarray}
ds^2 = -dt^2 + t^{2q} (dx^2 +dy^2) + t^{2 - 4q}dz^2 \, .
\end{eqnarray}
Here we have a stress-energy tensor which has all off-diagonal components 
vanishing:
\begin{eqnarray}
T_{ij} = \left ( \begin{array}{cccc} q\frac{(2-3 q)}{8 \pi t^2} & 0 &
0 & 0 \\ 0 & q\frac{(2-3 q)t^{2q}}{8 \pi t^2} & 0 & 0\\ 0 & 0 &
q\frac{(2-3 q)t^{2q}}{8 \pi t^2} & 0\\ 0 & 0 & 0 & q\frac{(2-3
q)t^{2-4q}}{8 \pi t^2}\end{array}\right ) \, .\
\end{eqnarray}

This metric forms a one parameter family of solutions of Einstein's
equations with a perfect stiff fluid. The parameter $q$ is related to
the energy density, as is obvious from the last equation. The
qualitative features of the expansion depend on $q$ in the following
way: for $q > 1/2$ the universe expands from a ``cigar'' singularity;
for $q = 1/2$, the universe expands purely transversally from an
initial ``barrel'' singularity; for $0 < q < 1/2$ the initial
singularity is ``point-like'' and if $q \leq 0$ we have a ``pancake''
singularity. The case $q=1/3$ corresponds to an isotropic universe
with a stiff fluid; the case $q=0$ is a region of Minkowski spacetime
in non-Cartesian coordinates. This family of metrics is ``Kasner-like''
in the sense that the sum of the exponents is equal to one, but the
sum of the squares is not equal to one except in the cases when $q=0$
or $q=2/3$, when we have the vacuum case.
 
\begin{figure}
\epsfxsize=3.0in
\epsfysize=3.0in
\epsfbox{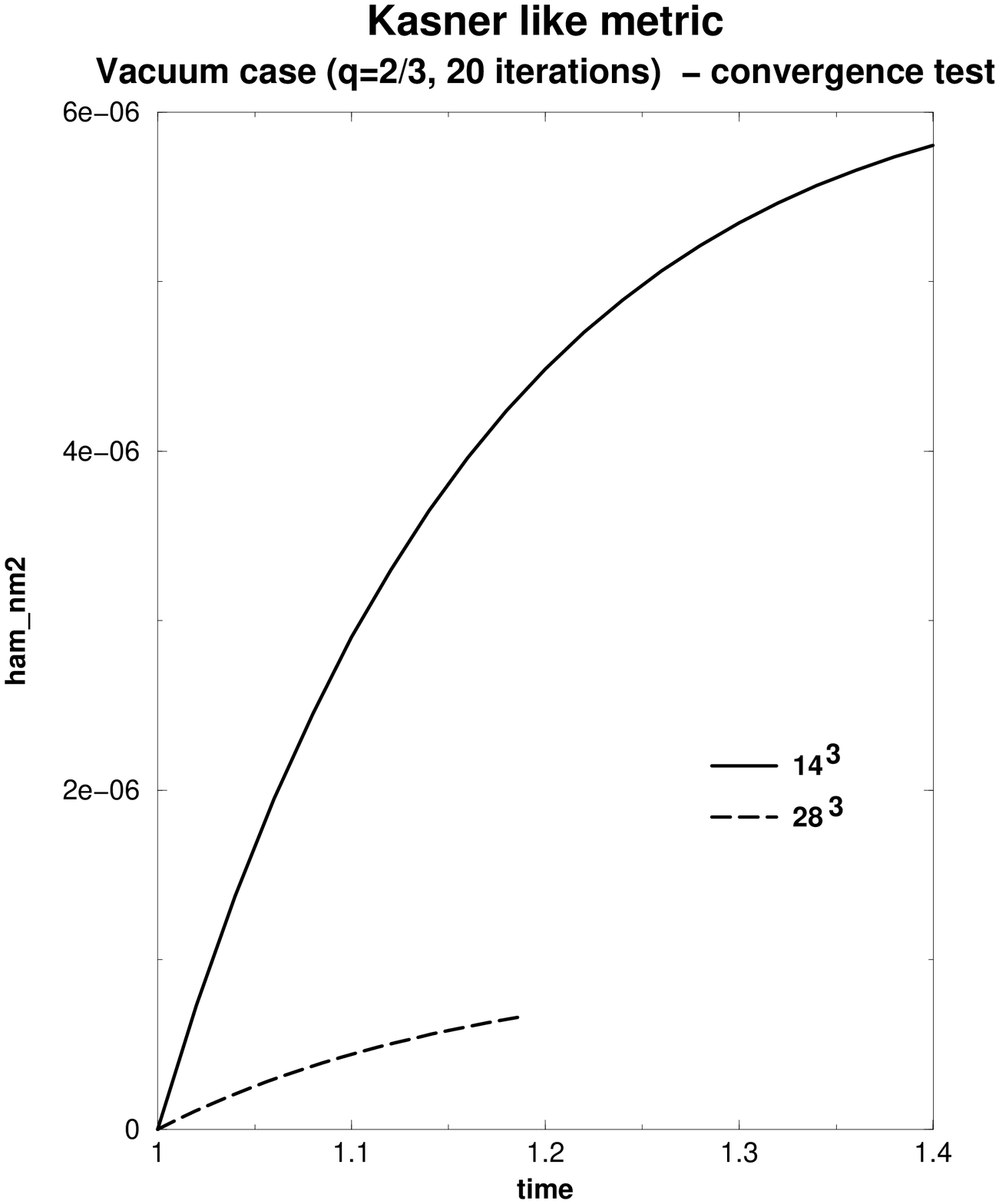}
\vspace{-3.0in}
\hspace{3.2in}
\epsfxsize=3.0in
\epsfysize=3.0in
\epsfbox{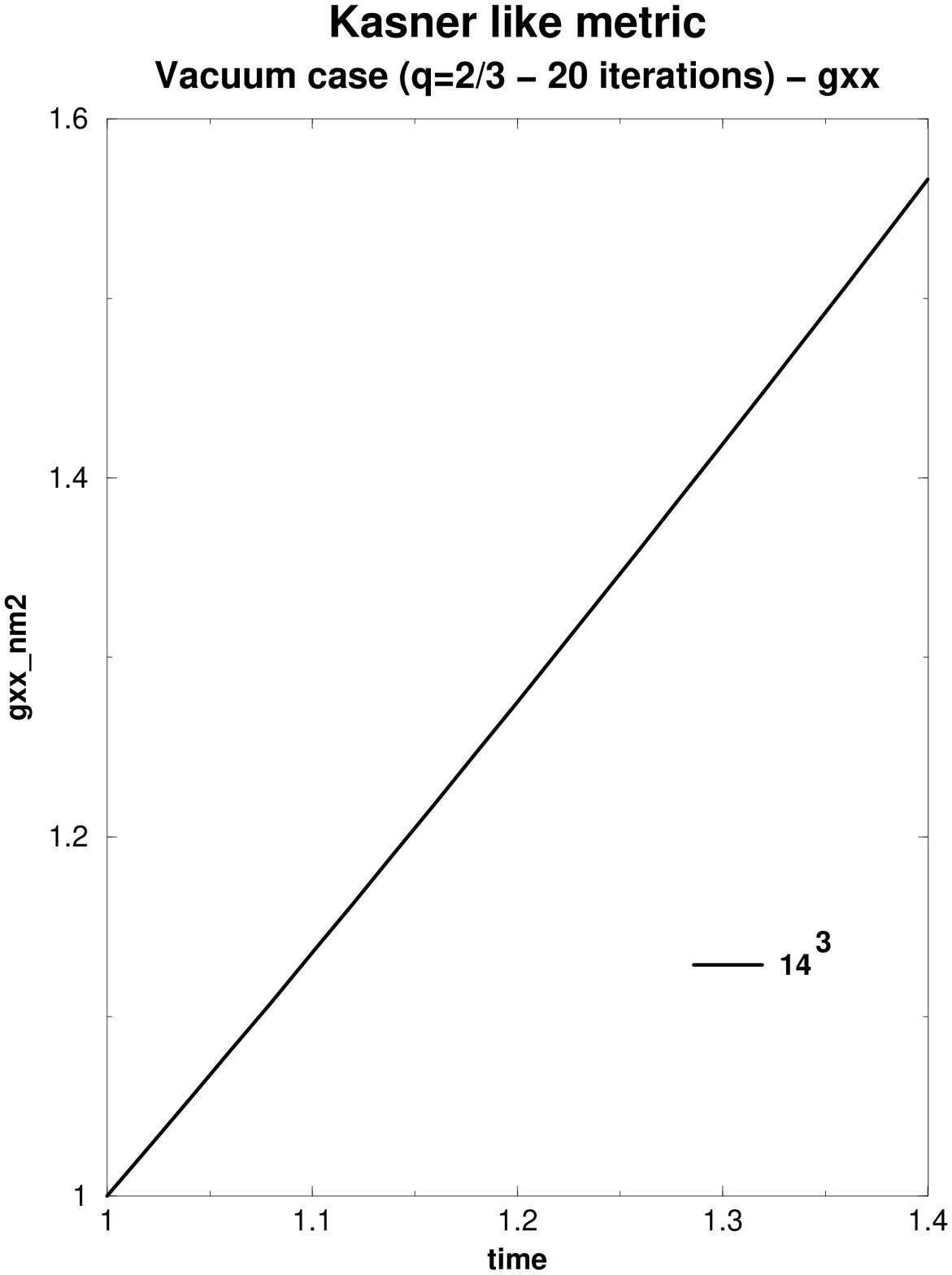}
\caption{Convergence of the Hamiltonian constraint (left panel) and 
  evolution of the L2 norm of the $g_{xx}$ metric component (right
  panel) for the Kasner-like spacetime - vacuum case ($q=2/3$) and 20
  iterations.}
\label{fig:kasner066}
\end{figure}

\begin{figure}
\epsfxsize=3.0in
\epsfysize=3.0in
\epsfbox{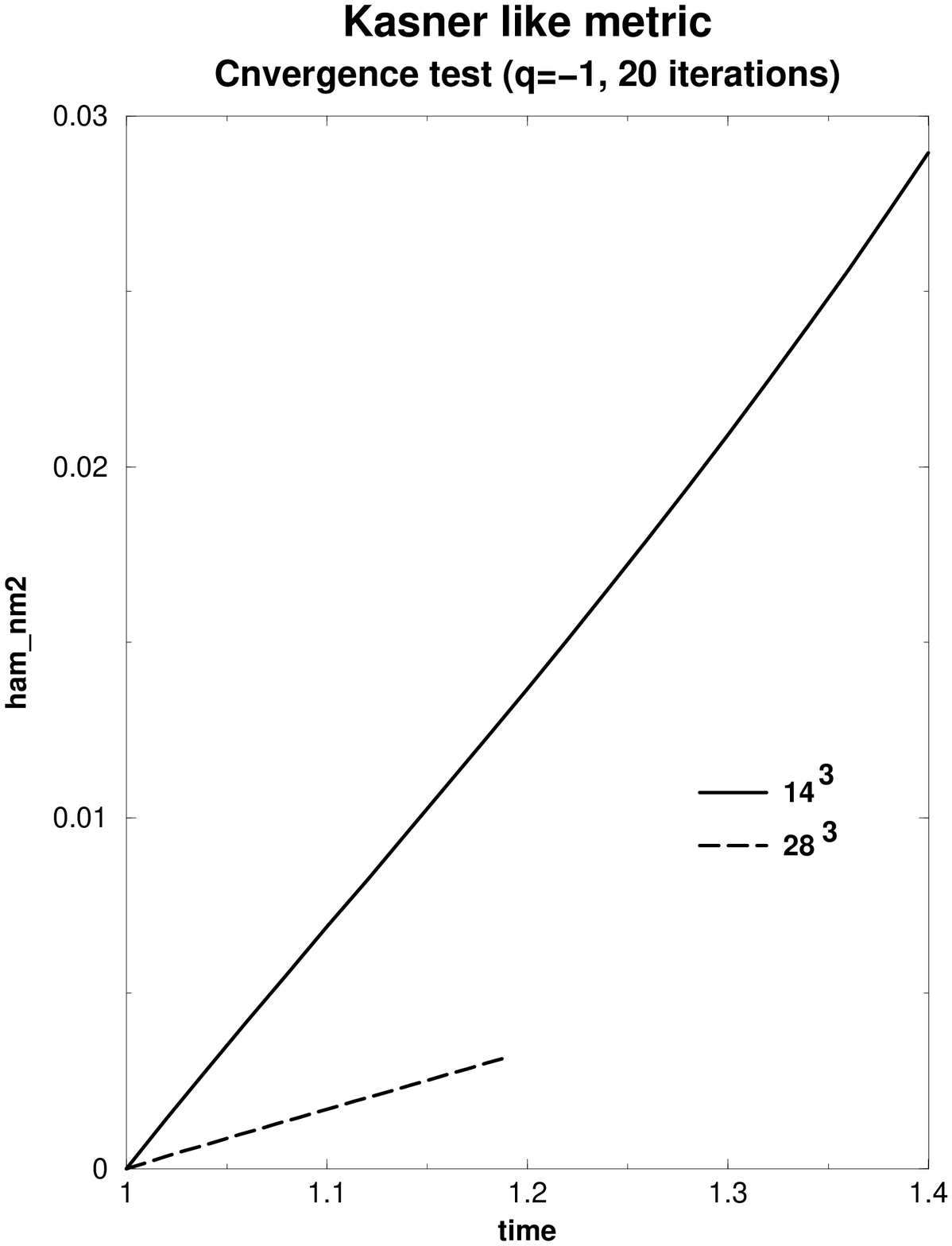}
\vspace{-3.0in}
\hspace{3.2in}
\epsfxsize=3.0in
\epsfysize=3.0in
\epsfbox{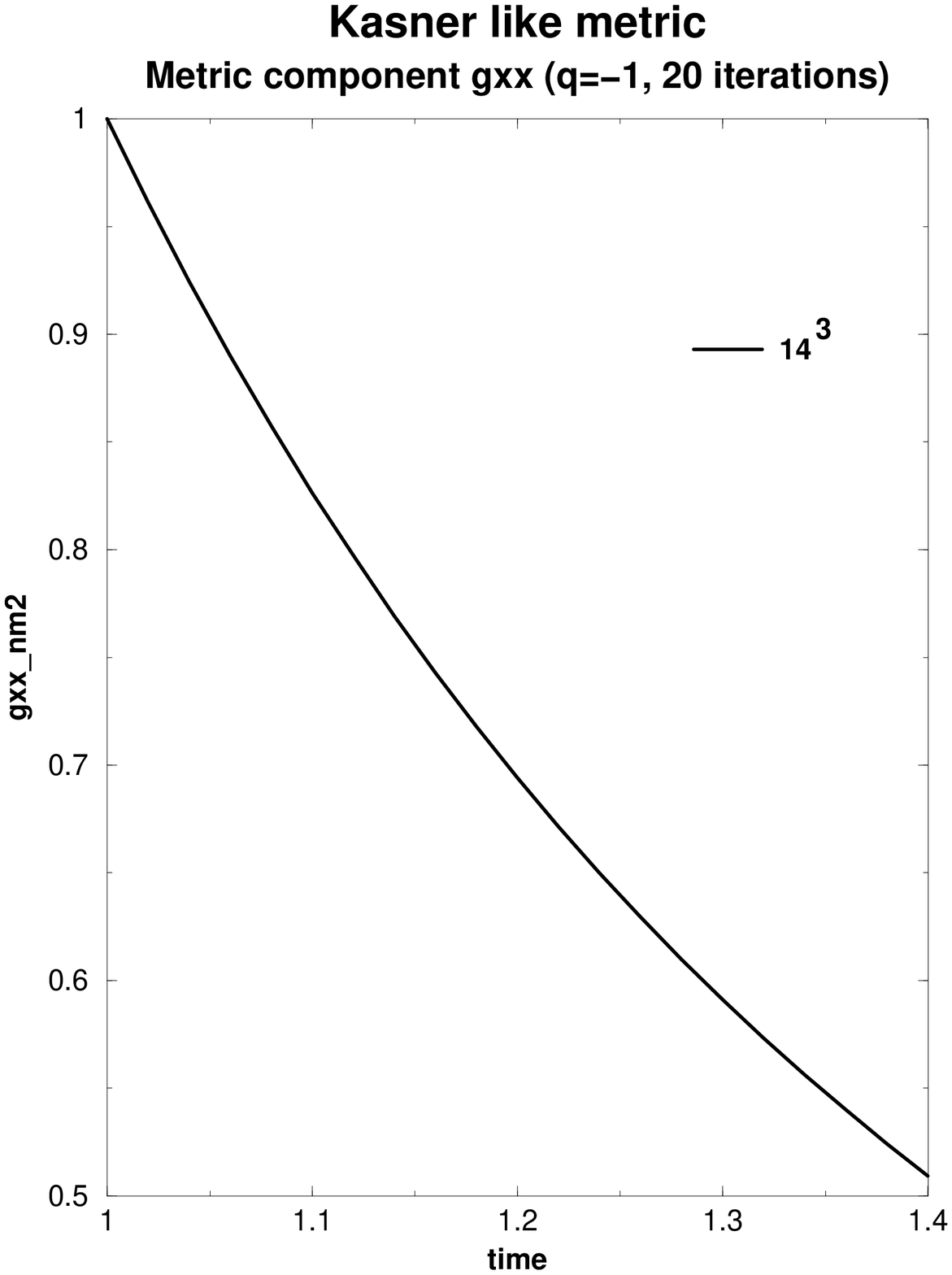}
\caption{Convergence of the Hamiltonian constraint (left panel) and 
  evolution of the L2 norm of the $g_{xx}$ metric component (right
  panel) for the Kasner-like spacetime for $q=-1$ and 20
  iterations.}
\label{fig:kasner-1}
\end{figure}

We have investigated the numerical behavior of this metric for
different values of the parameter $q$, looking specially at the
convergence of the Hamiltonian constraint, starting from an initial
time $t=1$ (since the metric is singular at $t=0$), using ``flat''
boundary conditions and, as in the previous cases, for both 20 and
2000 iterations using the BSSN formulation.  All convergence tests
were done with two resolutions, one with $14^3$ and the second with
$28^3$ points on the grid.  Figure~\ref{fig:kasner066} displays our
results for the vacuum case ($q=2/3$) and 20 iterations. As can be
clearly seen the convergence of the Hamiltonian constraint is of
second order. In the figure we show also the evolution of the L2 norm
of the $g_{xx}$ metric component during that time. We show the same
plots in figure~\ref{fig:kasner-1}, but this time for the value
$q=-1$ of the Kasner parameter.

\begin{figure}
\hspace{0.4in}
\epsfxsize=3.0in
\epsfysize=3.0in
\epsfbox{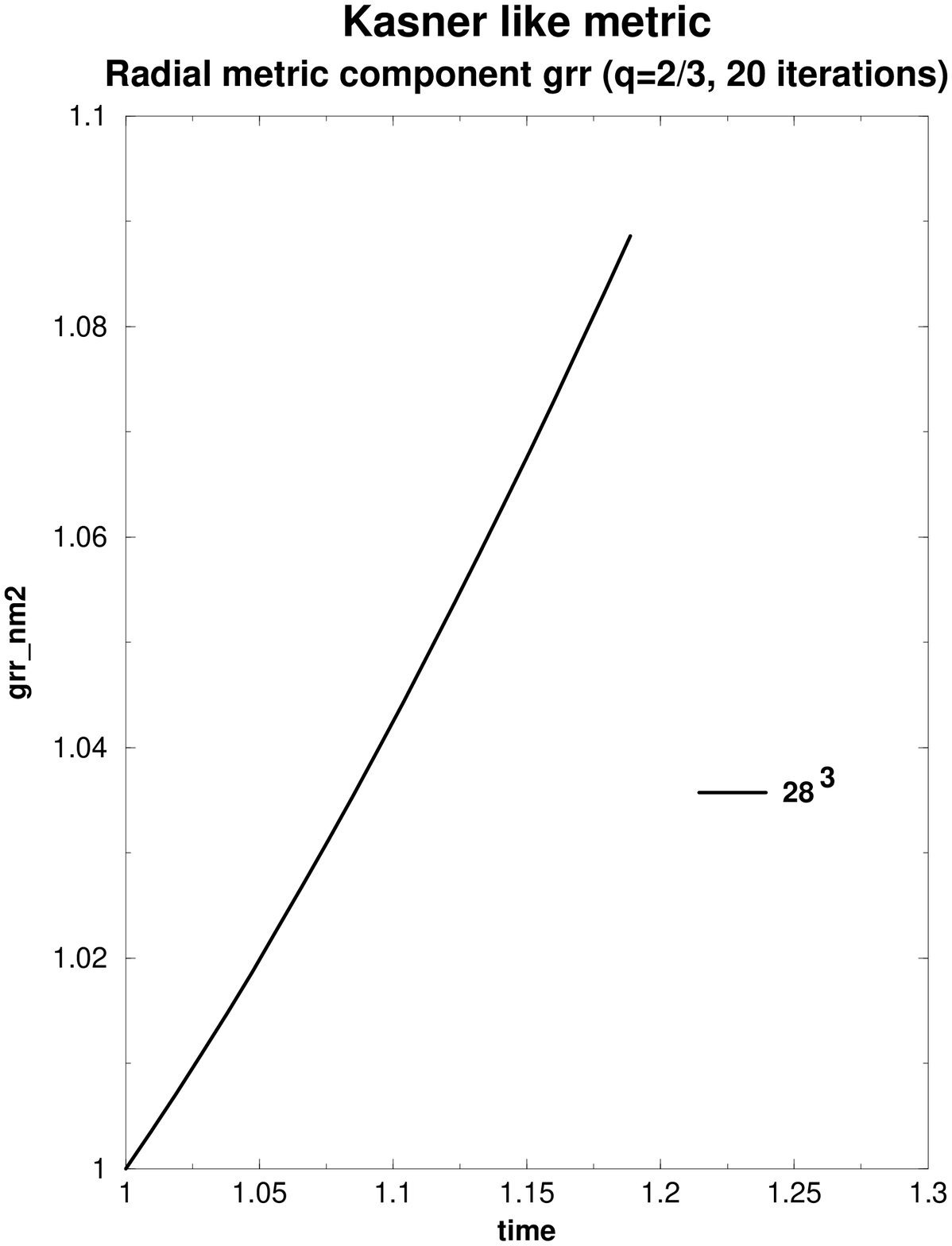}
\vspace{-3.0in}
\hspace{3.2in}
\epsfxsize=3.0in
\epsfysize=3.0in
\epsfbox{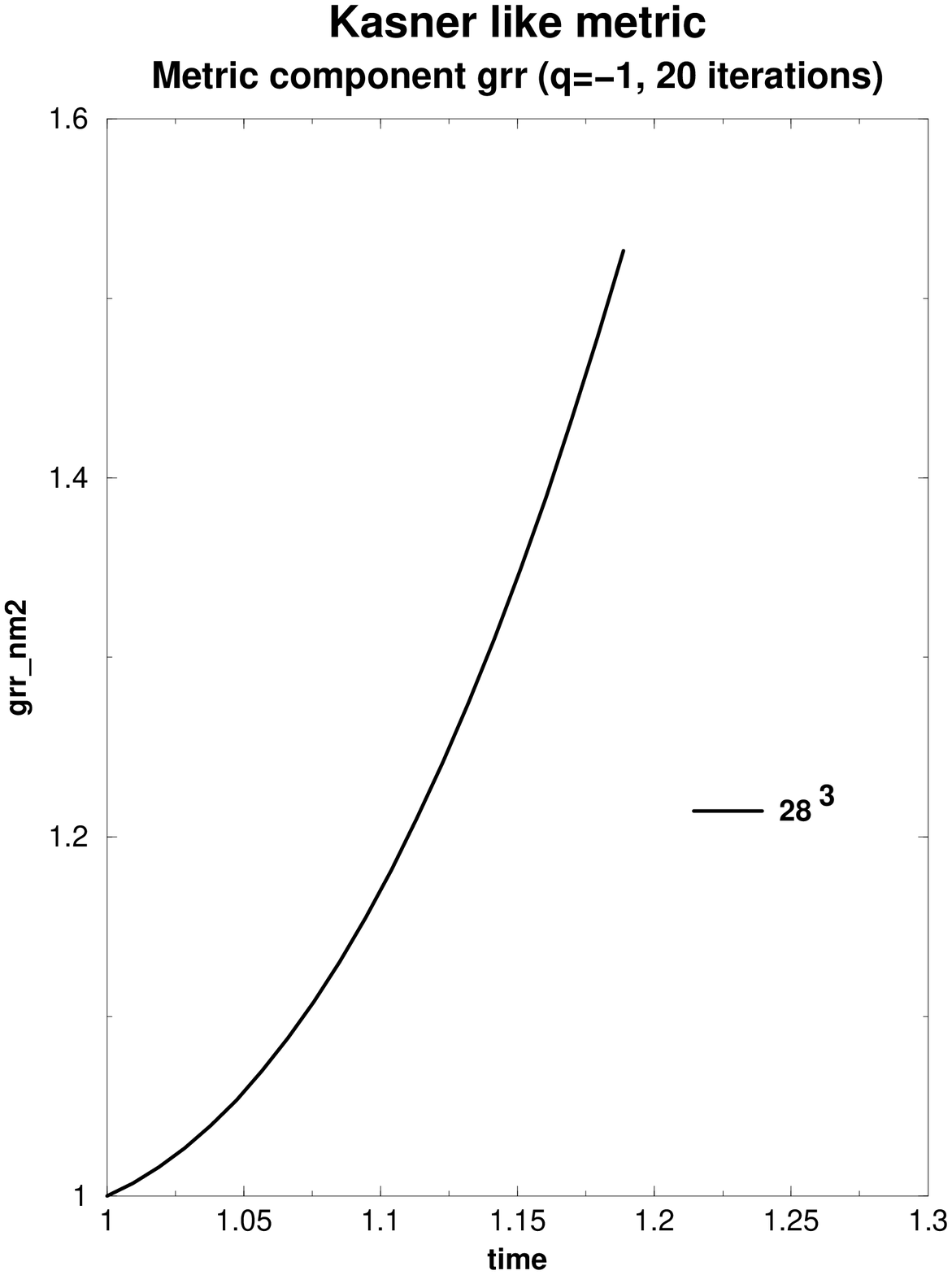}
\caption{Evolution of the L2 norm of the radial metric component
  $g_{rr}$ for the vacuum case ($q=2/3$, right panel) and for $q=-1$
  case (left panel) for the Kasner-like spacetime, with 20 iterations
  in both cases.}
\label{fig:kasner-sym}
\end{figure}

Figure~\ref{fig:kasner-sym} shows the evolution of the radial metric
component for the two previous cases, pointing out some differences
due to the special symmetry of this metric (see above and compare with
$g_{xx}$ from the previous figures).

\begin{figure}
\epsfxsize=3.0in
\epsfysize=3.0in
\epsfbox{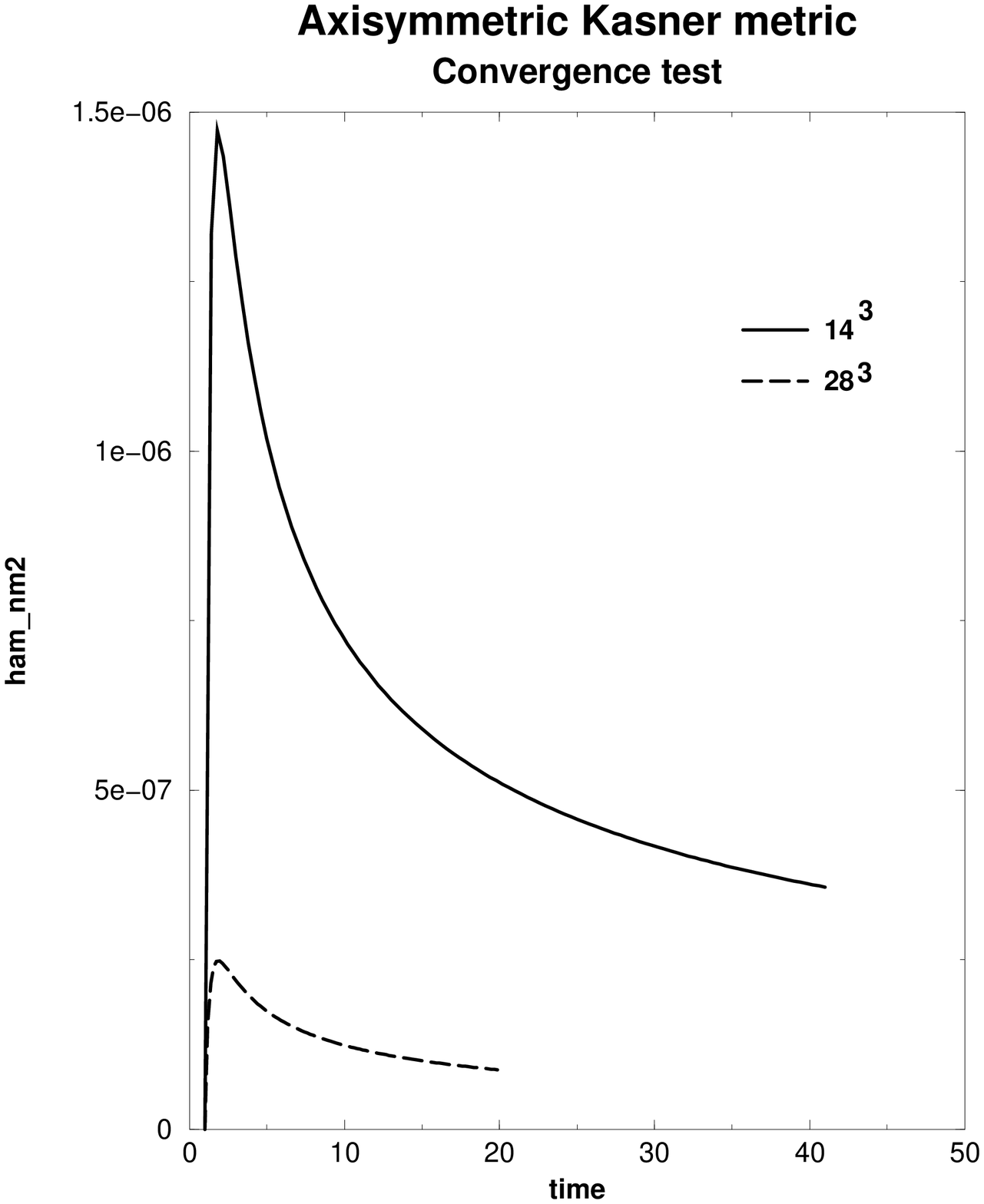}
\vspace{-3.0in}
\hspace{3.2in}
\epsfxsize=3.0in
\epsfysize=3.0in
\epsfbox{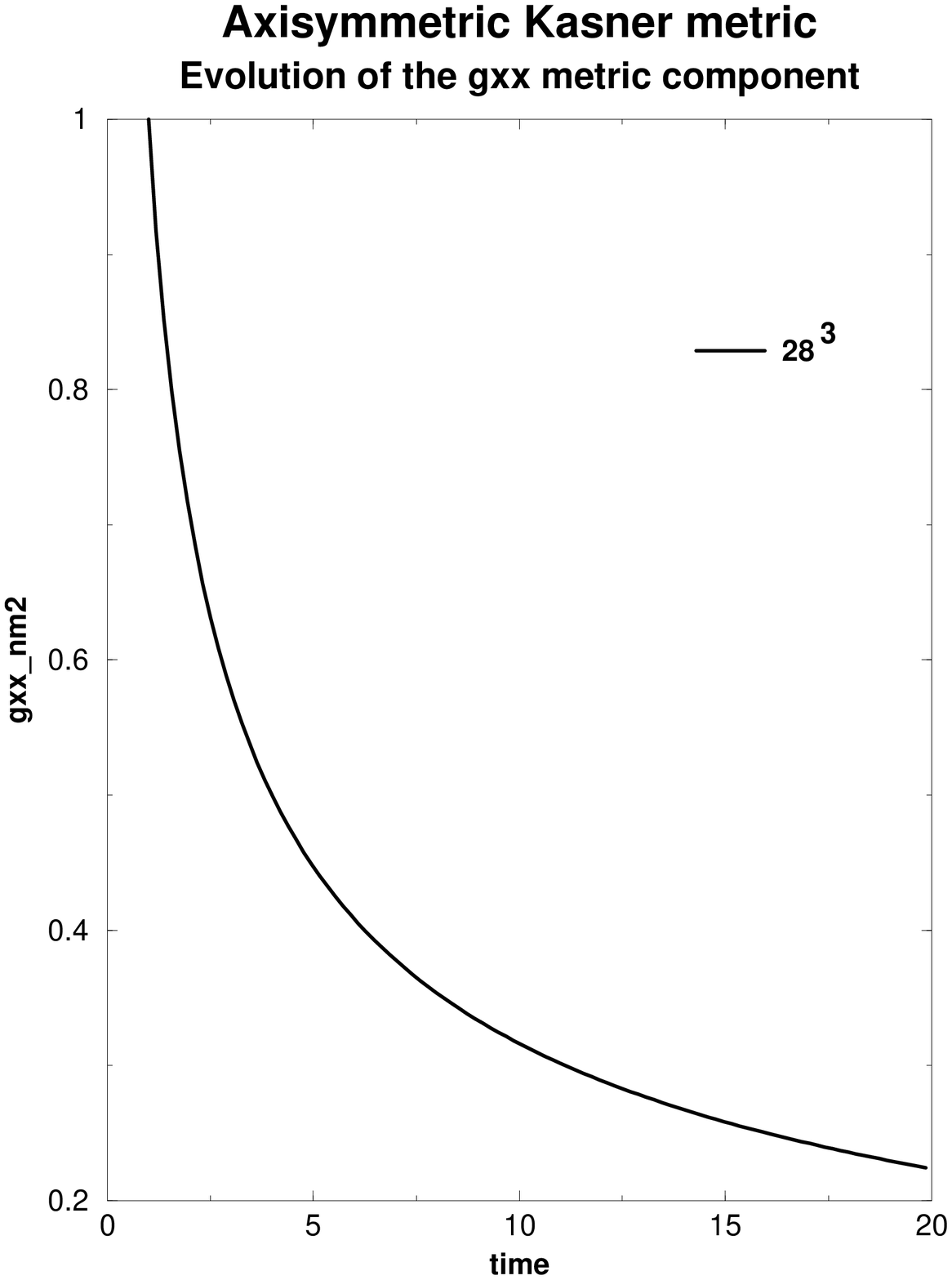}
\caption{Convergence of the Hamiltonian constraint (left panel) and
  the evolution of the L2 norm of $g_{xx}$ (right panel) for the
  axisymmetric Kasner spacetime using 2000 iterations.}
\label{fig:kasner-axi}
\end{figure}

Next we have considered another Kasner type metric, namely the Kasner 
axisymmetric spacetime (\cite{18}):
\begin{eqnarray}
ds^2 = -\frac{dt^2}{\sqrt{t}} + \frac{dx^2}{\sqrt{t}} + t dy^2
+ t dz^2 \, .
\end{eqnarray}
This is an exact solution of the vacuum Einstein equations, explicitly
homogeneous, and features a cosmological singularity at $t=0$.
Figure~\ref{fig:kasner-axi} shows some of our results for a ``long
time'' simulation (with 2000 iteration) showing the stability and
second order convergence of the code.

\begin{figure}
\epsfxsize=3.0in
\epsfysize=3.0in
\epsfbox{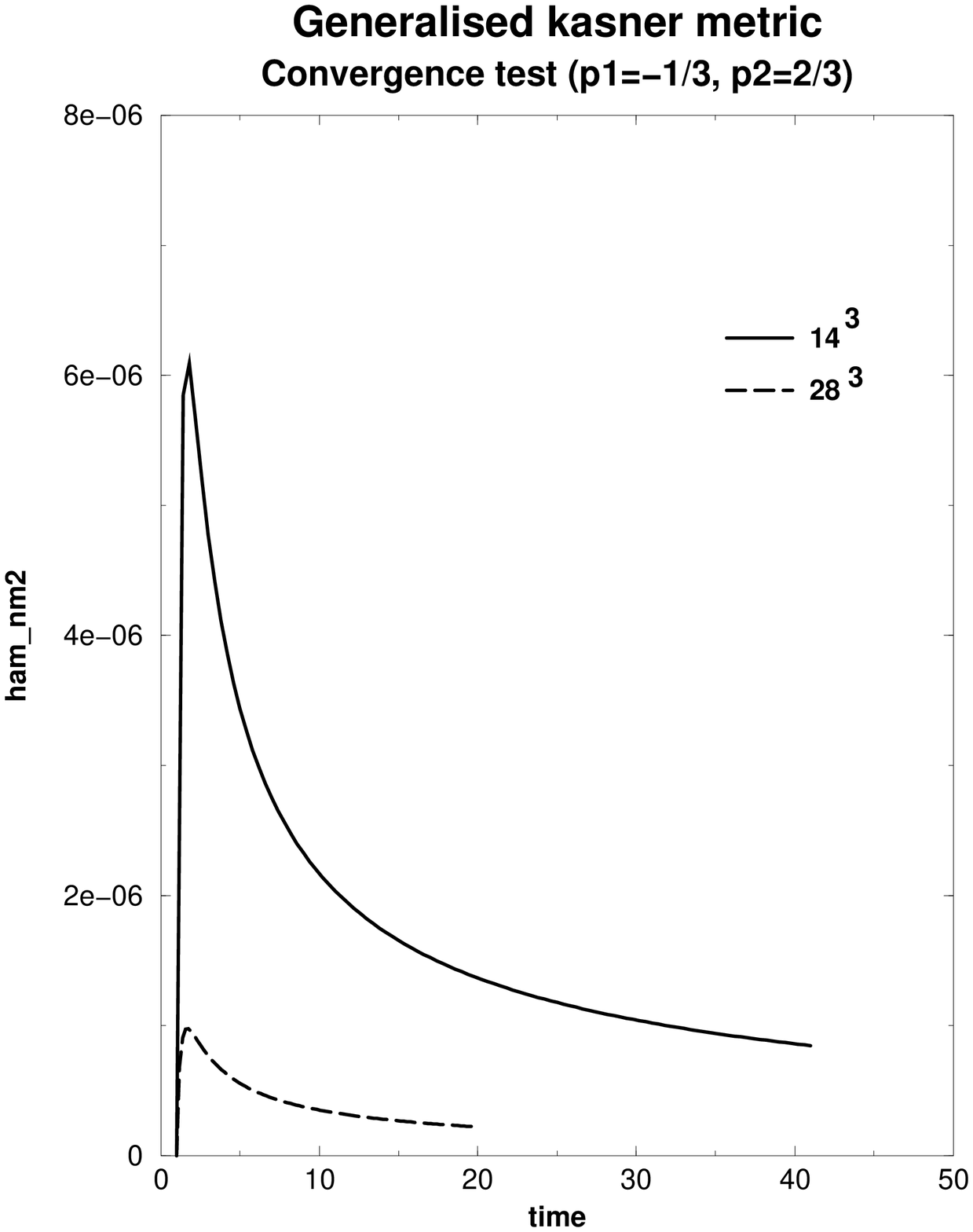}
\vspace{-3.0in}
\hspace{3.2in}
\epsfxsize=3.0in
\epsfysize=3.0in
\epsfbox{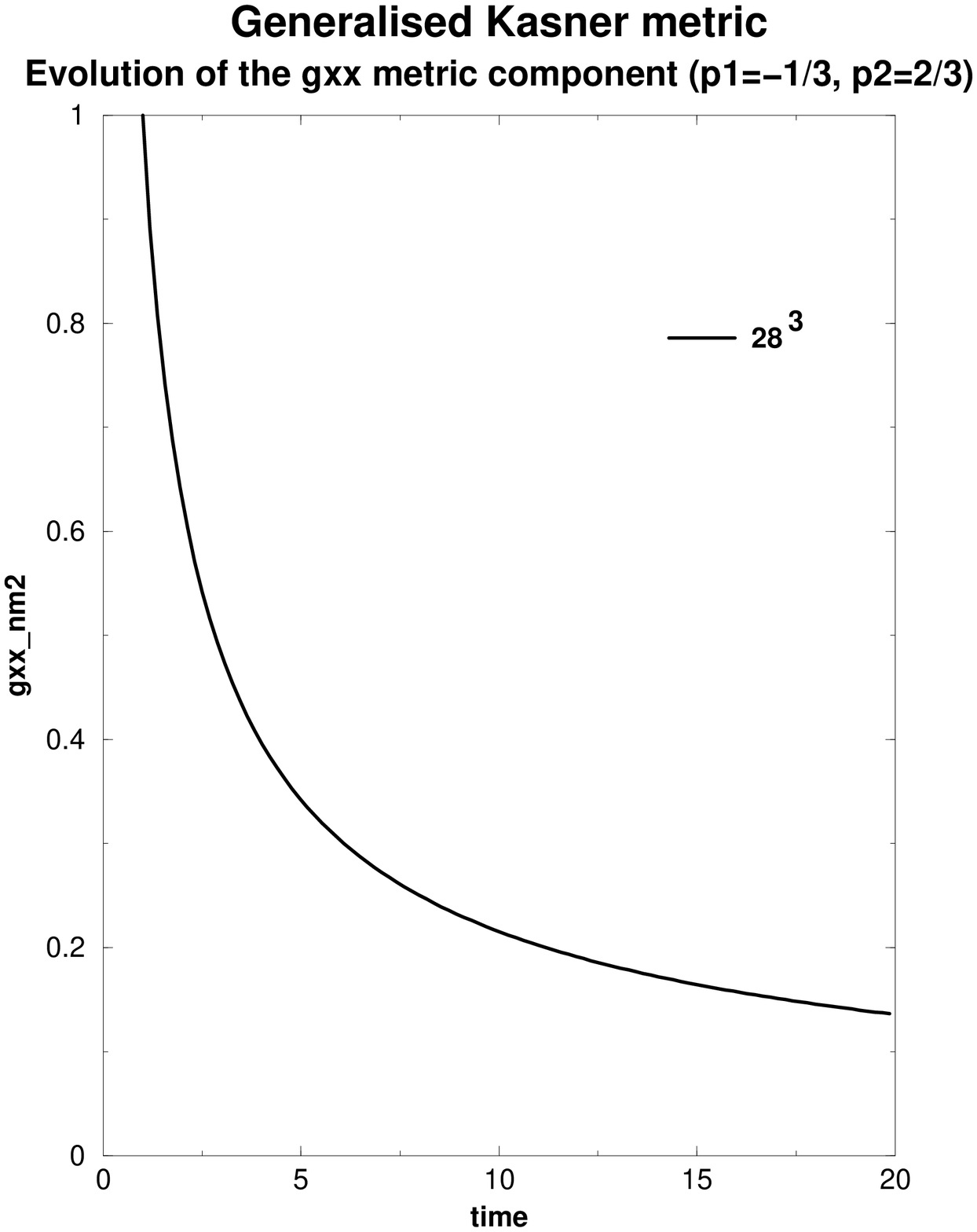}
\caption{Convergence of the Hamiltonian constraint (left panel) and 
  evolution of the L2 norm of $g_{xx}$ (right panel)
  for the generalized Kasner spacetime - $p_1=-1/3$, $p_2 = 2/3$, and
  2000 iterations.}\label{fig:kasner-gen}
\end{figure}

A generalization of the Kasner metrics can be introduced \cite{1}
with a metric of the form:
\begin{eqnarray}
ds^2 = -dt^2 +t^{2p_1}dx^2 + t^{2p_2}dy^2 + t^{2p_3}dz^2 \, ,
\end{eqnarray}
where the Kasner parameters $p_1$, $p_2$ and $p_3$ satisfy two relations:
\begin{eqnarray}
p_1+p_2+p_3=1 \hbox{~~~and~~~} p_1^2+p_2^2+p_3^2 = 1 \, .
\end{eqnarray}
Restricting ourselves only to two parameters, $p_1$ and $p_2$,
we have the following stress-energy tensor:
\begin{eqnarray}
T_{ij} = \left ( \begin{array}{cccc}
\frac{A}{8\pi t^2} & 0 & 0 & 0\\0 & \frac{A t^{2p_1-2}}{8 \pi} & 0 & 0\\
0 & 0 & \frac{A t^{2p_2-2}}{8\pi} & 0 \\
0 & 0 & 0 & \frac{A t^{-2p_1-2p_2}}{8 \pi}\end{array}\right ) \, ,
\end{eqnarray}
where $A = p_1 - p_1^2 +p_2 - p_2^2 - p_1 p_2$ (note the use of the above
first condition on the parameters, thus we have $p_3 = 1-p_1-p_2$).  
We have done several simulations for this metric. 
Figure~\ref{fig:kasner-gen} shows results for 2000
time iterations, using $p_1 = -1/3$, $p_2 = 2/3$.

\section{Robertson-Walker metric and other examples}

We have studied also other metrics, new or old ones form the Exact
thorn. We only mention here the Bianchi type I, G\" odel and 
Bertotti metrics.  A special case is the Kerr metric,
in two versions, one in Kerr-Schild coordinates and the other one in
standard Kerr (Cartesian) coordinates. Our results are similar to
those discussed above.

We have further studied the problem of introducing the
Robertson-Walker (\cite{16}) metric (\ref{RW})
as a generic case for the study of numerical cosmology.  Here we use
the scale factor of the universe $R(t)$ as a variable function through
the code, and two initial parameters: the initial mass-energy density 
and the initial scale factor of the universe. Because this case has 
several specific problems and features we shall report on it in a 
future article in preparation. This is intended to be done in 
connection with the further development of the Exact thorn code, where 
we shall introduce a scalar field coupled with gravitation in order to 
study inflationary models.

\section*{Acknowledgments}

D.N.V.  was supported, for one of his visits to
the AEI, by the DAAD through a NATO advanced research fellowship.
He is also  deeply indebted to Prof.
E.~Seidel for patience, continuous support and encouragement, making
possible his introduction to numerical relativity and to the Cactus code.

\end{document}